\documentstyle[12pt,fleqn]{article}

\newcommand{\abs}[1]{\hrule\par\begin{description}\item{Abstract: }
\it #1\par\end{description}\hrule\par\bigskip}
\newcommand{\pacs}[1]{\par\bigskip\noindent
PACS numbers:\par #1\par\bigskip}

\newcommand{\nn}{\nonumber}            
\newcommand{\ap}{\left.}               
\newcommand{\at}{\left(}               
\newcommand{\aq}{\left[}               
\newcommand{\ag}{\left\{}              
\newcommand{\cp}{\right.}              
\newcommand{\ct}{\right)}              
\newcommand{\cq}{\right]}              
\newcommand{\cg}{\right\}}             
\newcommand{\eq}[1]{\begin{equation}#1\end{equation}}  
\newcommand{\eqs}[1]{\begin{eqnarray}#1\end{eqnarray}}  

\newcommand{\ii}{\infty}                         
\newcommand{\fr}[2]{\mbox{$\frac{#1}{#2}$}}      
\newcommand{\Tr}{\,\mbox{Tr}\,}                  
\newcommand{\Res}{\,\mbox{Res}\,}                

\renewcommand{\Re}{\,\mbox{Re}\,}                
\newcommand{\lap}{\triangle}                     

\newcommand{\al}{\alpha}
\newcommand{\be}{\beta}
\newcommand{\ga}{\gamma}
\newcommand{\de}{\delta}

\newcommand{\ze}{\zeta}

\newcommand{\ka}{\kappa}
\newcommand{\la}{\lambda}
\newcommand{\ro}{\varrho}

\newcommand{\ta}{\tau}
\newcommand{\om}{\omega}

\newcommand{\te}{\vartheta}

\newcommand{\Ga}{\Gamma}

\newcommand{\Si}{\Sigma}
\newcommand{\Om}{\Omega}

\begin{document}

\title{Bose-Einstein condensation of scalar fields\\
on hyperbolic manifolds}

\author{Guido Cognola and Luciano Vanzo\\
\it Dipartimento di Fisica - Universit\`a di
Trento\thanks{Email: cognola@itnvax.cineca.it, vanzo@itncisca},
Italia and\\
\it Istituto Nazionale di Fisica Nucleare,
Gruppo Collegato di Trento}

\date{september 1992}

\maketitle

\abs{The problem of Bose-Einstein condensation
for a relativistic ideal gas on a {\rm 3+1} dimensional manifold
with a hyperbolic spatial part is analyzed in some detail.
The critical temperature is evaluated and its dependence of
curvature is pointed out.}


\section{Introduction}

Bose-Einstein condensation for a non relativistic ideal gas has a long
history\cite{eins24-22-261}. The physical phenomenon is well described in
many text books (see for example ref.\cite{huan63b}) and a rigorous
mathematical discussion of it was given by many
authors\cite{arak63-4-637,land79-70-43}.
The generalization to a relativistic idel Bose gas is non trivial and
only recently has been discussed in a series of
papers\cite{habe81-46-1497,habe82-23-1852,habe82-25-502}.

It is well known that in the thermodynamic limit
(infinite volume and fixed density) there is a phase transition of the
first kind in correspondence of the critical temperature at which the
condensation manifests itself. At that temperature, the first
derivative of some continuous thermodynamic quantities has a jump.
If the volume is keeped finite there is
no phase transition, nevertheless the phenomenon of condensation
still occurs, but the critical temperature in this case is not well
defined.

For manifolds with compact hyperbolic spatial part of the kind
$H^N/\Ga$, $\Ga$ being a discrete group of isometries for the
N-dimensional Lobachevsky space $H^N$,
zero temperature effects as well as finite temperature effects
induced by non-trivial topology, have been recently studied in some
detail
\cite{byts91-6-669,gonc91-8-L211,byts91-8-2269,byts92-9-1365,cogn92-33-222,byts9
   2-7-397,byts92r-255,cogn92r-258,byts92r-259}.
To our knowledge, a similar analysis has not yet been carried out
for non compact hyperbolic manifolds.

Hyperbolic spaces have remarkable properties.
For example, the continuous spectrum of the Laplace-Beltrami operator
has a gap determined by the curvature radius of $H^{N}$,
implying that massless fields have correlation
functions exponentially decreasing at infinity
(such a gap is not present for the Dirac operator).
For that reason, $H^{4}$ was recently proposed as an excellent infrared
regulator for massless quantum field theory and QCD\cite{call90-340-366}.
Critical behaviour is even more striking.
In two flat dimension vortex configurations of a complex
scalar field, the XY model for $He^{4}$ films have energy logarithmically
divergent with distance, while on $H^{2}$ it is finite.
This implies that the XY model is disordered
at any finite temperature on $H^{2}$.
Even quantum mechanics on $H^{2}$ has been the subject
of extensive investigations\cite{bala86-143-109}.
The manifold $H^{4}$ is also of interest as it is the Euclidean
section of anti-de Sitter space which emerges as the ground state
of extended supergravity theories.
The stress tensor on this manifold has been recently computed for
both boson and fermion fields using zeta-function
methods\cite{camp92-45-3591}.

In the present paper we shall discuss finite temperature effects and
in particular the Bose-Einstein condensation for a
relativistic ideal gas in a 3+1 dimensional ultrastatic space-time
${\cal M}=R\times H^3$.
We focus our attention just on $H^3$, because such a manifold could
be really relevant for cosmological and astrophysical applications.
To this aim we shall derive the thermodynamic potential for a charged
scalar field of mass $m$ on ${\cal M}$, using zeta function,
which on $H^3$ is exactly known.
We shall see that the thermodynamic potential has two branch points
when the chemical potential $\mu$ riches $\pm\om_o$, $\om_o^2=\ka+m^2$
being the lower bound of the spectrum of the operator $L_m=-\lap+m^2$
and $-\ka$ the negative constant curvature of $H^3$.
The values $\pm\om_o$ will be riched by $\mu=\mu(T)$ of course for
$T=0$, but also for $T=T_c>0$. This is the critical temperature at
which the Bose gas condensates.

The paper is organized as follows. In section 2 we study
the elementary properties of the Laplace-Beltrami operator on $H^3$;
in particular we derive its spectrum and build up from it
the related zeta-function. In section 3 we briefly recall how
zeta-function can be used in order to regularize the partition
function and we derive the regularized expression for the thermodynamic
potential. In section 4 we discuss the Bose-Einstein condensation and
derive the critical temperatures in both the cases of low and high
temperatures. In section 5 we consider in detail the low and high
temperature limits and derive the jump of the first derivative of the
specific heat. The paper end with some considerations on the
results obtained and some suggestions for further developments.

\section{The spectrum and the zeta function of
Laplace-Beltrami operator on $H^3$}

For the aims of the present paper, the 3-dimensional Lobachevsky
space $H^3$ can be seen as a Riemannian manifold of constant negative
curvature $-\ka$, with hyperbolic metric
$dl^2=d\ro^2+\sinh^2\!\ro(d\te^2+\sin^2\te\,d\te^2)$ and measure
$d\Om=\sinh^2\ro\,d\ro d\Si$, $d\Si$ being the measure on $S^2$.
For convenience, here we normalize the curvature $-\ka$ to $-1$.
In these coordinates, the Laplace-Beltrami operator $\lap$ reads

\begin{equation}
\lap=\frac{\partial^2}{\partial{\ro^2}}
+2\coth\ro\frac{\partial}{\partial\ro}
+\frac{1}{\sinh^2\ro}\lap_{S^2}
\end{equation}
$\lap_{S^2}$ being the Laplacian on the sphere.

The zeta function related to $-\lap$, can be easily computed if one
knows the density of states. For our aim, only states with
zero angular momentum will be important, so, in order to derive it,
it is sufficient to study radial wave functions of $-\lap$,
that is solution of equation

\begin{equation}
\frac{d^2u}{d\ro^2}+2\coth\frac{du}{d\ro}+\la u=0
\label{}
\end{equation}
which reduces to

\begin{equation}
\frac{d^2v}{d\ro^2}+(\la-1)v=0
\label{}
\end{equation}
with the substitution $u=v/\sinh\ro$. The bounded solution
of the latter equation is trivially given
by $v=A\sin\nu\ro$, $\nu$ being related to
the eigenvalue by $\la=\nu^2+1$. Such a solution remains bounded at
infinity provided $\nu$ stays real, or $\la\geq 1$.
Thus the continuum spectrum of
$-\lap$ on $H^3$ has a lower bound at $\la=1$
(or $\la=\ka$ in standard unities).
Note that in general this is not true on the compact
manifold $H^3/\Ga$, which has been considered in ref.\cite{cogn92r-258}.
We also notice that the wave operator propagates the field excitations
on the light cone, hence the gap should not be interpreted as a
physical mass.

For the radial solution normalized to 1 at the origin one gets

\begin{equation}
u_\nu(\ro)=\frac{\sin\nu\ro}{\nu\sinh\ro}
\label{}
\end{equation}
Now, the $L^2(d\Om)$ scalar product for $u_\nu(\ro)$ is

\begin{eqnarray}
(u_\nu,u_{\nu'})=\frac{4\pi}{\nu\nu'}
\int_{0}^{\ii}\sin\nu\ro\sin\nu'\ro\,d\ro
=\frac{2\pi^2}{\nu^2}\de(\nu-\nu')
\label{}
\end{eqnarray}
from which the density of states $\ro(\nu)=V\nu^2/2\pi^2$ directly
follows. As usual, we have introduced the large,
finite volume $V$ to avoid divergences. When possible, the limit
$V\to\ii$ shall be understood.

At this point the computation of zeta function is straightforward.
As we shall see in the following, what we are really interested in, is
the zeta function related to the operators $Q_{\pm}=L_m^{1/2}\pm\mu$.
The eigenvalues of $L_m$ are $\om^2(\nu)=\nu^2+a^2=\nu^2+\ka+m^2$,
then we get

\begin{eqnarray}
\ze(s;Q_{\pm})&=&\frac{V}{2\pi^2}\int_{0}^{\ii}
\at\om(\nu)\pm\mu\ct^{-s}\nu^2d\nu \nn \\
&=&\frac{s\Ga(s-3)V}{(4\pi)^{3/2}\Ga(s-1/2)(2a)^{s-3}}
F(s+1,s-3;s-\fr{1}{2};\fr{a\mp\mu}{2a})
\label{zetaQ}
\end{eqnarray}
where $F(\al,\be;\ga;z)$ is the hypergeometric function.
For its properties and its integral representations see
for example ref.\cite{grad80b}.
It has to be noted that eq.(\ref{zetaQ}) is the very same one has
on a flat space for a massive field with mass equal to $a$.
Here in fact, the curvature plays the role of an effective mass.

As we see from eq.(\ref{zetaQ}), the zeta function related to the
pseudo-differential operators $Q_\pm$ has simple poles at the points
$s_n=3,2,1,-1,-2,-3,\dots$ with residues
$b_n(\pm\mu)=\Res(\ze(s;Q_\pm),s_n)$ given by

\eq{
b_3(\pm\mu)=\frac{V}{2\pi^2};\\
b_2(\pm\mu)=\mp\frac{V\mu}{\pi^2};\\
b_1(\pm\mu)=\frac{V}{4\pi^2}(2\mu^2-a^2);
}

\eqs{
b_{-n}(\pm\mu)&=& c_{-n}
F(-n+1,-n-3;-n-1/2;\fr{a\mp\mu}{2a});\\
c_{-n}&=&
\frac{(-1)^nnV(2a)^{n+3}}{(4\pi)^{3/2}\Ga(-n-1/2)\Ga(n+4)};
\hspace{2cm}n\geq 1;\nn
}
Moreover, using the following property of the hypergeometric function:

\eq{
\frac{d}{dz}F(\al,\be;\ga;z)=\frac{\al\be}{\ga}F(\al+1,\be+1;\ga+1;z)
\label{Dhyper}
}
one easily sees that all $b_{-n}(\pm\mu)$ are odd or even polinomials
of $\mu$, according to whether $n$ is even or odd; this means that
$b_{-n}(\mu)=(-1)^{n+1}b_{-n}(-\mu)$ and the
function $\ze(Q_+)+\ze(Q_-)$ has simple poles at the points
$s=-2k+3$ with residues given by $2b_{-2k+3}(\mu)$ and
simple zeros for $s=-2k$ ($k=0,1,2,\dots$).
For the special values $\mu=\pm a$ we immediately get
$b_{-n}(a)=(-1)^{n+1}b_{-n}(-a)=c_{-n}$. Finally
from eq.(\ref{Dhyper}) we also derive the recurrence relations

\eq{
\partial_\mu b_{-n}(\pm\mu)
=\pm nb_{-(n-1)}
\label{Dmubn}
}

\section{The thermodynamic potential}

In order to define the grand canonical partition function
$Z(\be,\mu)=e^{-\be\Om(\be,\mu)}$, $\Om(\be,\mu)$ being the thermodynamic
potential, we suppose the charged scalar field to be in thermal
equilibrium
at finite temperature $T=1/\be$.
Then, according to ref.\cite{acto85-157-53}, one has

\begin{equation}
Z(\be,\mu)=\int_{\phi(0,\vec x)=\phi(\be,\vec x)}
{\cal D}\bar\phi{\cal D}\phi
e^{-\int_0^\be d\ta\int\bar\phi L_\mu\phi\sqrt{g}dV}
\label{Zbetamu}
\end{equation}
where $L_\mu=-(\partial_\ta-\mu)^2+L_m$ and the Wick rotation
$\ta=ix^0$ has to be understood.
In eq.(\ref{Zbetamu}) the integration has to
be taken over all fields $\phi(\ta,x^a)$ with
$\be$-periodicity with respect to $\ta$.

The eigenvalues of the whole operator $L_\mu$, say $\mu_{n,\nu}$ read

\begin{equation}
\mu_{n,\nu}=\left(\frac{2\pi n}{\be}+i\mu\right)^2+\om(\nu)^2
\\n=0,\pm 1,\pm 2,\dots
\end{equation}

Regularizing the partition function in eq.(\ref{Zbetamu})
in terms of zeta-function $\ze(s;L_\mu)$
related to the operator $L_\mu$ \cite{hawk77-55-133} one easily gets

\begin{eqnarray}
\Om(\be,\mu)&=&\frac{1}{\be}\log\det(\ell^{-2}L_\mu)=
-\fr{1}{\be}\ze'(0;\ell^{-2}L_\mu)\nn\\
&=&-\fr{1}{\be}[\log\ell^2\ze(0;L_\mu)+\ze'(0;L_\mu)]
\label{Omreg}
\end{eqnarray}
$\ell$ being an arbitrary normalization parameter
coming from the scalar path-integral measure. Note that $\ell$, which
has the dimensions of a mass, is necessary in order
to keep the zeta-function dimensionless for all $s$.
The finite temperature and $\mu$ dependent part of
the thermodynamic potential does not
suffer of the presence of such an arbitrary parameter.
On the contrary, $\ell$ enters in the regularized expression
of vacuum energy and this creates an ambiguity\cite{cogn92-33-222},
which is proportional to the heat kernel
expansion coefficient $K_N(L_m)$ related to $L_m$
(in general, $K_N(L_m)\neq 0$).
When the theory has a natural scale parameter, like the mass of the
particle or the constant curvature of the manifold, the ambiguity can
be removed by an "ad hoc" choice of $\ell$ \cite{dowk89-327-267}.

Here we would like to study the behaviour of thermodynamic quantities,
then we are only interested in the $\mu$ and $T$ dependent part of the
thermodynamic potential; that is a well defined quantity, which does not
need regularization. To compute it, it is not necessary to
use all the analytic properties of zeta function (for a careful
derivation of vacuum energy see for example
refs.\cite{cogn92-33-222,byts92r-255}).
Then we can proceed in a formal way and directly compute
$\log\det L_\mu$ disregarding the vacuum energy divergent term.

First of all we observe that

\begin{eqnarray}
\sum_{n=-\ii}^{\ii}\log(\om^2+(2\pi n/\be+i\mu)^2)
&=&\sum_{n=-\ii}^{\ii}\int\frac{d\om^2}{\om^2+(2\pi n/\be+i\mu)^2}\nn\\
&=&\int\frac{\be}{4\om}\left[\coth \fr{\be}{2}(\om+\mu)
+\coth \fr{\be}{2}(\om-\mu)\right]d\om^2 \\
&=&-\log\left[1-e^{-\be(\om+\mu)}\right]
-\log\left[1-e^{-\be(\om-\mu)}\right]-\be\om\nn
\end{eqnarray}
Using eq.(\ref{Omreg}), recalling that $\om^2=\nu^2+a^2$ and by
integrating over $\nu$ with the state density that we have derived in
the previous section, we get the standard result

\begin{eqnarray}
\Om(\be,\mu)&=&-\frac{1}{\be}\Tr\log L_\mu=E(\be,\mu)+E_v\nn\\
 &=&\frac{V}{2\pi^2\be}\int\ag
\log\left[1-e^{-\be(\om(\nu)+\mu)}\right]
+\log\left[1-e^{-\be(\om(\nu)-\mu)}\right]\cg\nu^2d\nu\\
&&+\frac{V}{2\pi^2}\int\om(\nu)\nu^2d\nu\nn
\label{logdetLmu}
\end{eqnarray}
The last term is divergent and represents the non regularized vacuum
energy that we throw away. The rest of the equation is exactly the $\mu$
and $T$ dependent part of the thermodynamic potential from
which we can derive all thermodynamic quantities.
$\Om(\be,\mu)$ as a function of the
complex parameter $\mu$, has branch points when $\mu^2$ is equal to
some $\om^2(\nu)$.
Thus the physical values of $\mu$ must be restricted by
$|\mu|\leq\om_o=a$. As we shall see in the next section,
the value $|\mu|=a$ can be riched for a critical temperature at which
Bose-Einstein condensation takes place.

In order to get low and high temperature expansions, it is convenient
to get different representations of $E(\be,\mu)$, which can be derived
in a rigorous way starting from eq.(\ref{Omreg}). For a compact
manifold $H^3/\Ga$ they have been derived in ref.\cite{cogn92r-258},
to which we refer the reader for more details. Here we just write down
the final results, observing that the expressions one has on $H^3$ are
the same one has on $H^3/\Ga$ when topological contributions are
dropped out. Then we have

\begin{equation}
E(\be,\mu)=-\frac{a^2V}{\pi^2}\sum_{n=1}^{\ii}\cosh n\be\mu
\frac{{\bf K}_2(an\be)}{n^2\be^2}
\label{Ebetamu}
\end{equation}

\begin{equation}
E(\be,\mu)=-\frac{1}{\pi i}\sum_{n=0}^{\ii}\frac{\mu^{2n}}{\Ga(2n+1)}
\int_{\Re s=c}\Ga(s+2n-1)\ze_R(s)\ze(\fr{s+2n-1}{2};L_m)\;\be^{-s}\;ds
\label{IREbetamu}
\end{equation}

\begin{equation}
E(\be,\mu)=-\frac{1}{2\pi i}\int_{\Re s=c}\Ga(s)\ze_R(s+1)
\aq\ze(s;Q_+)+\ze(s;Q_-)\cq\be^{-(s+1)}\;ds
\label{IRE}
\end{equation}
where ${\bf K}_2$ is the modified Bessel function,
$c$ is a sufficiently large real number and $\ze_R(s)$ is the usual
Riemann zeta-function.
The integral representations (\ref{IREbetamu}) and (\ref{IRE}),
which are valid for $|\mu|<a$, are useful for high temperature expansion.
On the contrary, the representation (\ref{Ebetamu}) in terms of modified
Bessel
functions is more useful for the low temperature expansion, since the
asymptotics of ${\bf K}_\nu$ is well known.

\section{Bose-Einstein condensation}

In order to discuss Bose-Einstein condensation we have to analyze the
behaviour of the charge density

\begin{equation}
\rho =z\frac{\partial\Om(V,\beta,z)}{V\partial z}\equiv f(z)-f(1/z)
\label{}
\end{equation}
in the infinite volume limit. Here

\begin{equation}
f(z)=\sum_{j}\frac{z}{V(\exp\beta\om_{j}-z)}
\label{}
\end{equation}
and the activity $z=\exp\beta\mu$ has been introduced. The $\om_{j}$ in the
sum are meant to be the Dirichlet eigenvalues for any normal domain
$V\subset H^{3}$. That is, $V$ is a smooth
connected submanifold of $H^{3}$ with non empty piecewise $C^{\infty}$
boundary. By the infinite volume limit we shall mean that a nested sequence
of normal domains $V_{k}$ has been choosen together with Dirichlet
boundary conditions and such that
$\bigcup V_{k}\equiv H^{3}$. The reason for this
choice is the following theorem due to Mac Kean
(see for example\cite{chav84b}):
\begin{description}
\item{---} if $\om_{ok}$ denotes
the smallest Dirichlet eigenvalue for any sequence of normal domains $V_{k}$
filling all of $H^{3}$ then $\om_{ok}\geq a$ and $\lim_{k\to\ii}\om_{ok}=a$.
(Although the above inequality is also true for Neumann boundary
conditions, the existence of the limit
in not assured to the authors knowledge).
\end{description}

Now we can show the convergence of the finite volume activity $z_{k}$ to
a limit point $\bar z$ as $k\rightarrow \infty$. To fix ideas, let us
suppose $\rho\geq 0$: then $z_{k}\in (1,\exp\be\om_{ok})$.
Since $\rho(V,\beta,z)$
is an increasing function of $z$ such that $\rho(V,\beta,1)=0$ and
$\rho(V,\beta,\infty)=\infty$, for each fixed $V_{k}$ there is a unique
$z_{k}(\bar\rho,\beta)\in (1,\be\exp\om_{ok})$ such that
$\bar\rho=\rho(V_{k},\beta,z_{k})$. By compactness, the sequence $z_{k}$
must have at least one fixed point $\bar z$ and as $\om_{ok}\to a^{2}$
as k goes to infinity, by Mc Kean theorem, $\bar z\in [1,\exp\beta a]$.

{}From this point on, the mathematical analysis of the infinite volume
limit exactly parallels the one in flat space for non relativistic
systems, as it is done in various references
\cite{ziff77-32-169,lewi74-36-1,land79-70-43}.
In particular, there is a critical temperature $T_c$ over which there are
no particles in the ground state.
$T_c$ is the unique solution of the equation

\eq{
\ro=\frac{\sinh\be a}{2\pi^2}\int_0^\ii
\frac{\nu^2d\nu}{\cosh\be a\sqrt{1+(\nu/a)^2}-\cosh\be a}
\label{Tc}
}

For $T>T_c$ we have $|\mu|<a$, the charge density of the
ground state $\ro_o=0$ and $\rho$ is given by

\begin{eqnarray}
\ro &=&\frac{\partial\Om(\be,\mu)}{V\partial\mu}
=\int_0^\ii\aq\frac{1}{e^{\be(\om(\nu)+\mu)}-1}
-\frac{1}{e^{\be(\om(\nu)-\mu)}-1}
\cq\frac{\nu^2}{2\pi^2}d\nu\nn\\
&=&\frac{\sinh\be\mu}{2\pi^2}\int_0^\ii
\frac{\nu^2d\nu}{\cosh\be\om(\nu)-\cosh\be\mu}
\label{robemu}
\end{eqnarray}
For $T\leq T_c$, $|\mu|$ remains
equal to $a$, the charge density $\ro_{ex}$ of the particles
in the excited states is given by eq.(\ref{Tc})
and the charge density of the particles in the ground
state is non vanishing and given by $\ro_o=\ro-\ro_{ex}$.
That is, below $T_c$ we have Bose-Einstein condensation.

So far we only have determined $\mu$ in the condensation region
where in fact it is equal to $a$. In the uncondensed region it is
determined by eq.(\ref{robemu}) as an implicit function of $\ro$ and $T$.
For massive particles,
it looks like the one, one has for the flat case\cite{habe81-46-1497},
with the only difference that the mass $m$ is replaced by
$a=\sqrt{\ka+m^2}$. The very difference between flat and hyperbolic
spaces occurs for massless particles. We shall return on this important
point in a moment.

Solutions of eq.(\ref{Tc}) can be easily obtained in the two cases
$\be a\gg 1$ and $\be a\ll 1$
(in the case of massive bosons these correspond to
non relativistic and ultrarelativistic limits respectively).
We have in fact

\eq{
\ro\simeq\frac{T^3}{2\pi^2}\int_0^\ii
\frac{Tx^2dx}{e^{x^2/2a}-1}
=\at\frac{aT}{2\pi}\ct^{3/2}\ze_R(3/2);
\\ \be a\gg 1
\label{ro1}
}

\eq{
\ro\simeq\frac{aT^2}{2\pi^2}\int_0^\ii
\frac{x^2dx}{\cosh x -1}
=\frac{aT^2}{3};
\\ \be a\ll 1
\label{ro2}}
from which we get the corresponding critical temperatures

\eq{
T_c=\frac{2\pi}{a}\at\frac{\ro}{\ze_R(3/2)}\ct^{2/3};
\\ \be a\gg 1
\label{Tc1}
}

\eq{
T_c=\at\frac{3\ro}{a}\ct^{1/2};
\\ \be a\ll 1
\label{Tc2}
}
and the corresponding charge densities of particles in the ground state

\eq{
\ro_o=\ro\aq 1-(T/T_c)^{3/2}\cq;
\\ \be a\gg 1
\label{ro01}
}

\eq{
\ro_o=\ro\aq 1-(T/T_c)^2\cq;
\\ \be a\ll 1
\label{ro02}
}
In the flat case, that is for $\ka=0$, eqs.(\ref{Tc1}) and (\ref{ro01})
reduce to the results given in many textbooks (see for example
ref.\cite{huan63b}), while eqs.(\ref{Tc2}) and (\ref{ro02})
agree with the ones given in ref.\cite{habe81-46-1497}.

It has to bo noted that for massless bosons, the condition
$|\mu|\leq a$ does not require $\mu=0$ like in the flat space,
because $a>0$ also for massless particles. This implies that the
critical temperature is always finite and so, unlike in the flat case,
$\ro$ is always different from $\ro_o$. As has been noticed
in ref.\cite{habe81-46-1497}, on a flat manifold,
the net charge of massless bosons resides in the Bose-Einstein
condensed ground state. This never happens if the spatial manifold is
$H^3$. In fact, because of curvature,
massive and massless bosons have a similar behaviour.

As it is well known, at the critical temperature, continuous
thermodynamic quantities may have a discontinuous derivative
(first order phase transition). This derive from the fact that
the second derivative of the chemical potential $\mu$ is
discontinuos for $T=T_c$.
Of course, $\mu'=d\mu/dT=0$ for $T<T_c$, since in this region $\mu$
is constant. By considering the constant charge density $\ro$ as a
function of temperature and taking the total derivative with respect to $T$ of
eq.(\ref{robemu}), we see that

\eq{
\mu'=-\frac{\partial_T\ro(T,\mu)}
{\partial_\mu\ro(T,\mu)}
\label{Dmu}
}
and since ${\partial_\mu\ro}$ diverges for $\mu=a$ we obtain
$\mu'(T^+_c)=0$. This is not the case of $\mu''$. In fact we shall
see that $\mu''(T^+_c)$ is different from zero and therefore $\mu''(T)$ is a
discontinuous function of temperature.
This implay that the first derivative of the
specific heat $C_V$ has a jump for $T=T_c$ given by

\eq{
\ap\frac{dC_V}{dT}\right|_{T_c^+}
-\ap\frac{dC_V}{dT}\right|_{T_c^-}
=\ap\mu''(T_c^+)\frac{\partial U(T,\mu)}{V\partial\mu}\right|_{T=T_c^+}
=-\ap T_c^+\mu''(T_c^+)
\frac{\partial\ro(T,\mu)}{\partial T}\right|_{T=T_c^+}
\label{jumpDTCV}
}
$U(T,\mu)$ being the internal energy,
which can be derived by means of equation

\eq{
U(\be,\mu)=-\mu\ro V+\frac{\partial}{\partial\be}
\be\Om(\be,\mu)
\label{Ubemu}
}

To go further in the computation, it is convenient to integrate
eq.(\ref{IREbetamu}) in order to obtain a series representation of
$E(\be,\mu)$ valid for $T>T_c$.
This is the topic of the next section.

\section{Low and high temperature expansions}

The low temperature expansion can be easily obtained by using
eq.(\ref{Ebetamu}) and recalling that for real values of $s$
the asymptotics for the modified Bessel function reads

\begin{equation}
{\bf K}_2(s)\sim\sqrt{\frac{\pi}{2s}}e^{-s}
\sum_{k=0}^{\ii}\frac{\Ga(k+5/2)}{\Ga(k+1)\Ga(-k+5/2)}(2s)^{-k}
\label{}
\end{equation}
Then, for small $T$ we have

\eq{
E(\be,\mu)\simeq -\frac{a^4V}{(2\pi)^{3/2}}
\sum_{n=1}^{\ii}e^{-n\be(a-|\mu|)}\at\frac{1}{an\be}\ct^{5/2}
\sum_{k=0}^{\ii}\frac{\Ga(k+5/2)}{\Ga(k+1)\Ga(-k+5/2)}
\at 2an\be\ct^{-k}
\label{lowex}
}
and deriving with respect to $\mu$
\eq{
\ro(T,\mu)\simeq -\at\frac{aT}{2\pi}\ct^{3/2}
\sum_{n=1}^{\ii}\frac{e^{-n\be(a-|\mu|)}}{n^{3/2}}
\label{rolow}
}
where only the leading terms have been taken into account.
Setting $|\mu|=a$ we again obtain the charge density (\ref{ro1}).
By deriving eq.(\ref{rolow}) we obtain

\eq{
\mu''(T_c^+)=-\frac{2C^2}{TA^2}; \\
T\frac{\partial \ro}{\partial T}=\frac{3}{2}\ro
}
where $A=2.363$ and $C=-2.612$ are two coefficients of the expansion

\eq{
\sum_{n=1}^{\ii}\frac{e^{-nx}}{n^{5/2}}=Ax^{3/2}+B+Cx+Dx^2+...
}
Now, using eq.(\ref{jumpDTCV}), we have the standard result

\eq{
\ap\frac{dC_V}{dT}\right|_{T_c^+}
-\ap\frac{dC_V}{dT}\right|_{T_c^-}
=\frac{3\ro C^2}{T_c A^2}=\frac{3.66\ro}{T_c}
}

The high temperature expansion could be obtained by using
eq.(\ref{IREbetamu}), like in ref.\cite{cogn92r-258}.
Here we shall use eq.(\ref{IRE}), because
for the aim of the present paper it is more convenient.
Using the properties of $\ze(s;Q_\pm)$, which we have discussed in
section 2, we see that the integrand function in eq.(\ref{IRE})

\begin{equation}
\ze_R(s+1)\Ga(s)[\ze(s;Q_+)+\ze(s;Q_-)]\;\be^{-(s+1)}
\end{equation}
has simple poles at $s=3,1,0,-3,-5,-7,\dots$ and a double pole at
$s=-1$.
Integrating this function on a closed path
containing all the poles, we get the high temperature expansion,
valid for $T>T_c$  (here $\ga$ is the Euler-Mascheroni constant)

\eqs{
E(\be,\mu)&\simeq&-V\aq
\frac{\pi^2}{45\be^4}-\frac{1}{12\be^2}(a^2-2\mu^2)
+\frac{(a^2-\mu^2)^{3/2}}{6\pi\be}\cp\nn\\
&&+\ap\frac{\mu^2}{24\pi^2}(3a^2-\mu^2)
+\frac{a^4}{16\pi^2}\left(\log \fr{a\be}{4\pi}+\ga-\fr{3}{4}\right)
\cq\label{highex}\\
&&-2\sum_{n=1}^{\ii}\frac{\ze_R(2n+1)b_{-(2n+1)}\be^{2n}}
{(-2\pi)^n(2n+1)}\nn
}
where we have used the formula

\eq{
\ze'_R(-2n)=\frac{\Ga(2n+1)\ze_R(2n+1)}{(-2\pi)^n}
}

Deriving eq.(\ref{highex}) with respect to $\mu$ and using
eq.(\ref{Dmubn}) we get the expansion for the charge density

\eqs{
\ro&\simeq&-\frac{\mu T^2}{3}+\frac{\mu T(a^2-\mu^2)^{1/2}}{2\pi}
-\frac{\mu(3a^2-2\mu^2)}{12\pi^2}\label{rohigh}\\
&&-\frac{2}{V}\sum_{n=1}^{\ii}
\frac{\ze_R(2n+1)b_{-2n}T^{-2n}}{(-2\pi)^n}\nn
}
For $\mu=a$, the leading term of this expression gives again
the result(\ref{ro2}).

{}From eq.(\ref{rohigh}), by a strightforward computation
and taking only the leading terms into account, one gets

\eq{
\mu''(T^+_c)\simeq -\frac{12\pi^2}{9a};
\\ \ap T\partial_T\ro(T,\mu)\right|_{T=T^+_c}\simeq -2\ro
}
and finally, from eqs.(\ref{jumpDTCV}) and (\ref{Ubemu})

\eq{
\left.\frac{dC_V}{dT}\right|_{T_c^+}
-\left.\frac{dC_V}{dT}\right|_{T_c^-}
\simeq-\frac{32\ro\pi^2}{9a}
}
in agreement with the result of ref.\cite{habe81-46-1497}.

\section{Conclusion}

We have shown that both massless and massive scalar fields on the
Lobachevsky space $H^{3}$ exhibites Bose-Einstein condensation at a
critical temperature depending on the curvature of the space:
the higher is the curvature radius the higher is the critical temperature.
The treatment is not intended to be complete in any sense but we were
able to display a curvature effect on thermodynamic quantities at the
most elementary level. In particular, due to curvature, massless
charged bosons have a finite $T_{c}$ in contrast to the flat space
result. The difference can be traced back to the existence of a gap
in the spectrum of the Laplace operator on $H^{3}$. We also pointed out
that the infinite volume limit is under good control only for Dirichlet
boundary conditions, for which the smallest eigenvalue has its limit
value precisely at the gap of the continuous spectrum.

The manifold $R\times H^{3}$ with the product metric
is not even a solution of Einstein equations nor it
has constant sectional curvature. In fact, the unique simply connected
static solution of Einstein equations with hyperbolic spatial part
is anti-de Sitter space-time for which there is a non trivial $g_{00}$
component of the metric tensor representing a constant gravitational
field. Then the equilibrium temperature is position dependent and a
precise discussion requires the elucidation of the notion of local
equilibrium, as has been stressed in ref.\cite{ziff77-32-169}.
The same remark can be applied to any static gravitational field.
More generally, one can
ask whether there are stationary geometries, i.e. a space-time with a
Killing vector field with non-vanishing vorticity, such that the
spatial 3-geometry is locally hyperbolic. This would be the gravitational
counterpart of an ideal Bose gas in a rotating vessel, for which there are
critical velocities at which correspond the appearance of quantized
vortices\cite{blat55-100-476}.

However, it has to be noted that a Robertson-Walker solution of
Einstein equations, together with the assumption that the material
content of the universe expands adiabatically,
can represent a manifold of the form we have considered.
The problem we have studied then can find physical applications
in the standard model of the universe.

\end{document}